\documentclass{article}

\usepackage{microtype}
\usepackage{graphicx}
\usepackage{subfigure}
\usepackage{booktabs} 

\usepackage{hyperref}

\usepackage{amsmath}
\usepackage{amsfonts}

\usepackage[accepted]{icml2019}

\icmltitlerunning{Risk-Sensitive Compact Decision Trees for Autonomous Execution in Presence of Simulated Market Response}

\newcommand{\E}{\operatorname{E}}

\newcommand{\vct}[1]{\boldsymbol{#1}}
\newcommand{\mtx}[1]{\boldsymbol{#1}}

\newcommand{\set}[1]{\mathcal{#1}}

\newcommand{\argmin}[1]{\underset{#1}{\operatorname{arg}\,\operatorname{min}}\;} 
\newcommand{\argmax}[1]{\underset{#1}{\operatorname{arg}\,\operatorname{max}}\;} 

\newcommand{\vb}{\vct{b}}

\newcommand{\vr}{\vct{r}}

\newcommand{\vw}{\vct{w}}

\newcommand{\vphi}{\vct{\phi}}

\newcommand{\mA}{\mtx{A}}

\newcommand{\mPhi}{\mtx{\Phi}}

\newcommand{\setA}{\set{A}}

\newcommand{\setC}{\set{C}}

\newcommand{\setS}{\set{S}}

\begin{document}

\twocolumn[
\icmltitle{ Risk-Sensitive Compact Decision Trees for Autonomous Execution in Presence of Simulated Market Response
             }



\icmlsetsymbol{equal}{*}

\begin{icmlauthorlist}
\icmlauthor{Svitlana Vyetrenko}{qr_jpm}\renewcommand{\thefootnote}{\fnsymbol{footnote}}\footnotemark
\icmlauthor{      Shaojie Xu}{ece_gatech}
\end{icmlauthorlist}

\icmlaffiliation{qr_jpm}{Quantitative Research, J. P. Morgan Chase, New York, NY, USA}
\icmlaffiliation{ece_gatech}{Department of Electrical and Computer Engineering,  Georgia Institute of Technology, Atlanta, GA, USA}

\icmlcorrespondingauthor{Svitlana Vyetrenko}{svitlana.s.vyetrenko@jpmorgan.com}

\icmlkeywords{Reinforcement learning, q-learning, optimized execution, limit order book simulator, decision tree, market impact, ICML}

\vskip 0.3in
]



\printAffiliationsAndNotice{}
\renewcommand{\thefootnote}{\fnsymbol{footnote}}\footnotetext{Opinions and estimates constitute our judgement as of the date of this paper, are for informational purposes only and are subject to change without notice. This paper is not the product of J.P. Morgan and therefore, has not been prepared in accordance with legal requirements to promote the independence of market research, including but not limited to, the prohibition on the dealing ahead of the dissemination of investment research. This paper is not intended as market research, a recommendation, advice, offer or solicitation for the purchase or sale of any financial product or service, or to be used in any way for evaluating the merits of participating in any transaction. It is not a market research report and is not intended as such. Past performance is not indicative of future results. Please consult your own advisors regarding legal, tax, accounting or any other aspects including suitability implications for your particular circumstances. J.P. Morgan disclaims any responsibility or liability whatsoever for the quality, accuracy or completeness of the information herein, and for any reliance on, or use of this material in any way.}

\begin{abstract}
We demonstrate an application of risk-sensitive reinforcement learning to optimizing execution in limit order book markets. We represent taking order execution decisions based on limit order book knowledge by a Markov Decision Process; and train a trading agent in a market simulator, which emulates multi-agent interaction by synthesizing market response to our agent's execution decisions from historical data. Due to market impact, executing high volume orders can incur significant cost. We learn trading signals from market microstructure in presence of simulated market response and derive explainable decision-tree-based execution policies using risk-sensitive Q-learning to minimize execution cost subject to constraints on cost variance. 
\end{abstract}

\section{Introduction}
\label{introduction}

\subsection{Problem statement}
\label{problem statement}
\begin{figure}[ht]
\begin{center}
\centerline{\includegraphics[scale=.45]{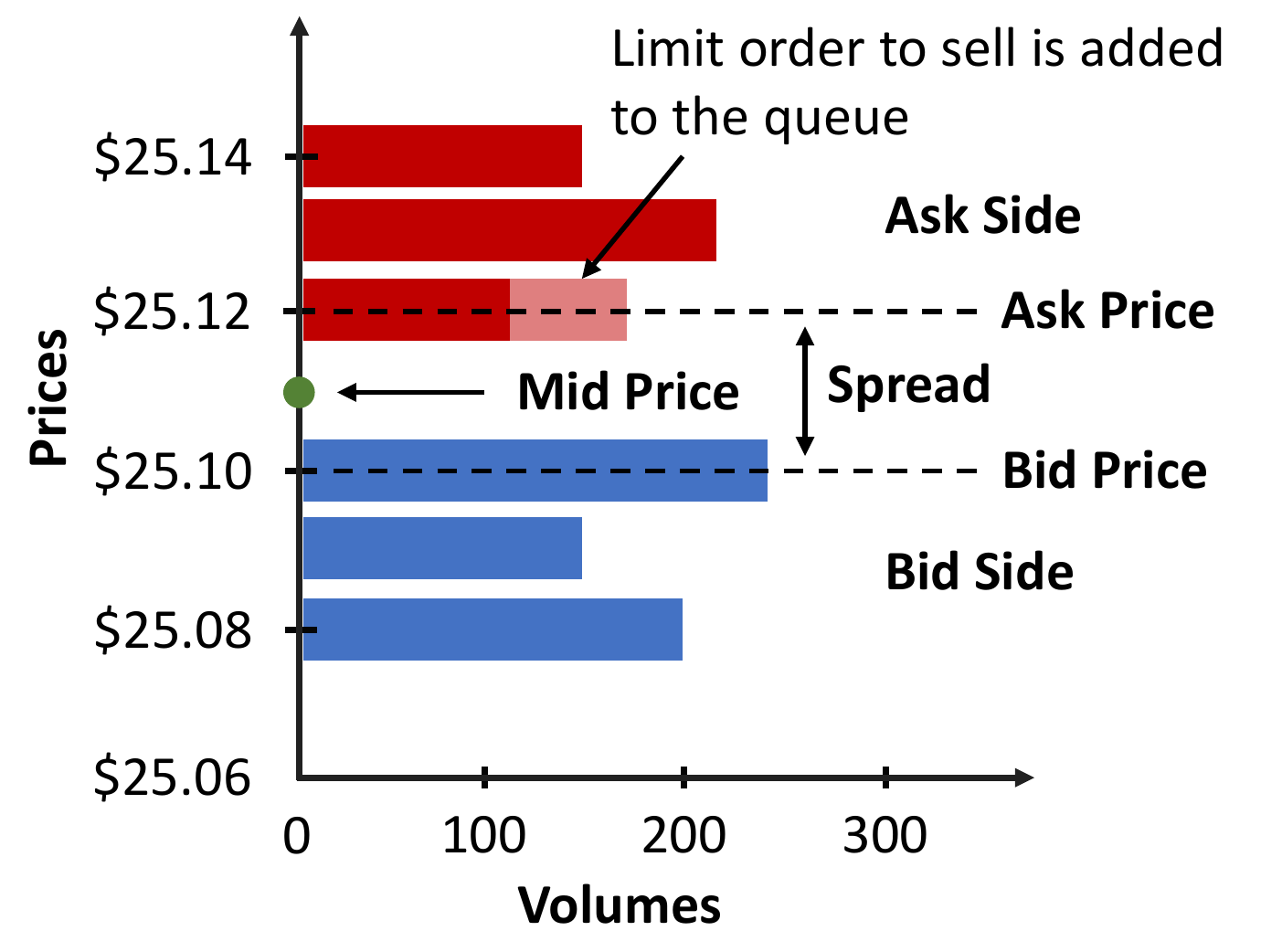}}
\vspace{-.5cm}
\caption{Visualization of the LOB structure.}
\label{fig:LOB}
\end{center}
\vspace{-1cm}
\end{figure}

Increasingly large market volumes are traded today electronically across multiple asset classes. Electronic limit order book (LOB) is the list of orders that is maintained by a trading venue to indicate ``buy'' and ``sell'' interest of market participants. A trading venue also maintains a matching engine to determine which orders can be filled. A market order targets immediate consumption of available liquidity at the opposite level at the cost of paying the spread. A limit order queues a resting order in the LOB at the side of the book of the market participant \cite{Bouchaud02}. Hence, placing a limit order will incur no spread cost. It will, however, incur risk of not being filled if the market moves away while the limit order is waiting to be matched with the opposite interest in the queue.  A chart that visualizes LOB structure is shown in Figure \ref{fig:LOB}.

We are addressing the problem of minimizing risk-adjusted cost of executing high volume orders in LOB markets. Executing large market orders at once will cause adverse moves in price that is called market impact, therefore, it is a common practice for market participants to break large parent orders into smaller
child orders that are then executed in the market \cite{Almgren00, Cartea15}. Hence, a market participant that is interested in minimizing risk-adjusted cost is facing the following questions:
\begin{enumerate}
\item How to split a large parent order into small child orders and at what times to execute child orders?
\item Is it better to execute child orders passively (i.e. via limit orders) or aggressively (i.e. via market orders) in presence of other market participants?
\end{enumerate}

To answer the above questions, we use model-free risk-sensitive reinforcement learning as a tool for sequential decision-making that allows us to model market response to our agent's actions. We model taking order execution decisions based on signals derived from LOB knowledge by a Markov Decision Process; and train an execution agent in a LOB simulator, which simulates multi-agent interaction. In practice we are also interested in interpretability of our agent's decisions. As a result, we construct explainable execution policies that are given by decision trees.

\subsection{Related work}
\label{Related work}

In traditional execution approaches (such as TWAP and VWAP),  the parent order execution schedule is known ahead of the first child order placement. In the TWAP (Time-Weighted Average Price) strategy, the parent order is split into child orders evenly over a specified time period. In the VWAP (Volume-Weighted Average Price) strategy, volumes of child orders are allocated proportionally to the observed or projected market volumes over a given time period.  Minimizing total expected cost of execution plus constant $\lambda>0$ times the variance of the cost was introduced in \cite{Almgren00} as another approach to static execution scheduling under assumptions of linear market impact and independent price returns. Choice of constant $\lambda$ determines risk sensitivity of the execution strategy, which is aligned with traders' intuition about risk aversion.

The developments in Markov decision processes (MDP) and reinforcement learning (RL) provide new frameworks for decision making in dynamic systems. In these frameworks, a dynamic system is characterized by several states. An agent makes a certain action given its current state, and the dynamics is encoded as the transition probability of moving from one state to another given the action. In financial trading applications, the statistics of a system is often unknown and difficult to model. In such cases, repeating the process a number of times in a simulated environment allows to eliminate the need of knowing transition probabilities explicitly, and the optimal policy can be learned from the gained simulated experience. The above-described model-free RL approach received attention in financial trading literature \cite{Neuneier97, moody2001learning, Nevmyvaka06, Spooner18}. Algorithms such as TD-learning \cite{sutton1988learning}, Q-learning \cite{watkins1989learning, watkins1992q}, SARSA \cite{rummery1994line}, and policy gradient \cite{williams1992simple, sutton2000policy} were applied to train an agent to adapt dynamically to the changing market conditions without modeling the dynamics of the trading environment explicitly. 

When both state space and action space are discrete and small, optimal policies can be learned in a tabular form. As a response to rapid electronification of trading, \cite{Nevmyvaka06} presented a first large-scale empirical application of RL for optimized trade execution that built upon tabular Q-learning. For continuous state spaces, function approximation is often used. 
\cite{Spooner18} used a linear combination of tile codings as the value function approximation to derive an RL market-making strategy. \cite{moody2001learning} proposed trading strategies designed by policy-gradient-based RL algorithm. \cite{Neuneier97} proposed a Q-learning based algorithm to dynamically adjust the allocation between two assets based on market and portfolio statistics. In the above work, the Q-values are approximated by a neural network and are updated using semi-gradient methods.

When designing an RL-based trading system, it is important to consider the interpretability of a learned trading strategy. As the learned policy is essentially a function mapping from state space to action space, it can be represented by a decision tree \cite{quinlan1986induction, safavian1991survey}. Such representations are simple and explainable.  \cite{Uther98} shows how to construct decision-tree-based discretization for continuous state space RL, where the algorithm alternates between learning the MDP dynamics and splitting states into distinctive children states. Controlling the number of state variables can also improve the policy interpretability. Batch RL feature selection and feature learning methods \cite{Liu15, Kolter09, Geist12} can improve our understanding of signals that contribute to autonomous trading agent construction.

Most of MDP and RL research focuses on maximizing the expected rewards, however, the variance of the received rewards is also of much interest in financial applications due to its representation of risk. The variance of the cumulative reward comes from both the stochastic reward received at each time step and the MDP dynamics. Selecting a risk-averse policy in an MDP was studied in \cite{chung1987discounted,filar1989variance,mannor2011mean}, but it has been difficult to transfer these algorithms to RL. \cite{bertsekas1995dynamic} suggested using augmented state space, introducing an auxiliary state variable that keeps track of the cumulative past reward. This approach, however, can significantly increase the state space and lead to inefficient learning algorithms. Assignment of immediate rewards based on utility functions was used in \cite{Ritter17} and is aligned with risk aversion theories in finance and economics. However, it requires making assumptions about the reward variance and effectively discourages choosing actions that receive immediate rewards with large variance. 

A framework that  suggests to apply a parameterized risk-sensitive transformation function to the temporal difference was presented in \cite{Mihatsch02}. This approach does not require making additional assumptions about the variance and can be effectively incorporated into existing RL algorithms. Moreover, changing the risk-sensitivity parameter in the transformation function provides flexibility to choose from a wide range of policies based on our risk preference - which resonates with the well-established execution approach given in \cite{Almgren00}. We apply this framework to our RL-based execution strategy design.

\subsection{Our contributions}
\label{Our contributions}

The main contribution of this paper is establishing a model-free risk-sensitive reinforcement learning framework for deriving compact decision tree policies for algorithmic execution. We highlight that training  a reinforcement learning agent in simulated market environment purely on historical data does not take  into account interaction of the agent with the market. In academic literature, the assumption of negligible market impact of aggressive trading if the size of agent orders is small and sufficient amount of time is allowed between consecutive aggressive trades is typically made (e.g. \cite{Spooner18}).  We relax these assumptions by introducing a simple method to synthesize market impact of aggressive trades from historical data. The proposed method essentially models our trading agent's interaction with other market participants, hence making the simulated LOB environment and reinforcement learning procedure more realistic.

Using a risk-sensitive Q-learning procedure, we can train execution strategies with a required level of risk aversion \cite{Mihatsch02}. Note that by doing so, we do not need to make any assumptions on the reward variance.

The choice to represent an execution agent by a decision tree is motivated by increased interpretability of decision tree policies.  In practice, we need to explicitly understand trading signals and be able to explain agent's risk. Historical LOB data is typically noisy, therefore, training an agent with the inherently smaller number of degrees of freedom allows to potentially prevent overfitting.

We outline the following steps in building our framework for decision tree execution agent training:
\begin{enumerate}
\item \textbf{Simulator:} Build LOB simulator capable of synthesizing the market impact of aggressive trading (Section~\ref{simulator}).
\item \textbf{Risk-sensitive Q-learning:} Derive decision-tree policies via risk-sensitive tabular Q-learning (decision tree boundaries are inferred from learned tabular policies) (Sections ~\ref{agent training} and ~\ref{decision_tree_example}).
\item \textbf{Feature selection via least-squares policy iteration:} Use linear parametric architecture to select a small number of statistically significant features given a large number of preselected input features (Section~\ref{feature selection}).
\end{enumerate}

\section{Simulated market environment}
\label{simulator}

\subsection{Simulator assumptions}
\label{assumptions}

We have constructed a market simulator that replays the LOB using prices and volumes at multiple LOB levels and completed trade data provided by the exchange (all of the above information is available at historical times). Note that historical information about queue position of resting orders and about queue cancellations is unknown. Both in training and in real-time trading, times of placement of new orders into LOB are determined by an autonomous trading agent. When an agent decides to place a new passive order into the LOB, it is assumed to be placed at the back of the queue. Upon observations about historical changes in LOB volumes at every time step, we make assumptions about our queue order position since we have no information about queue cancellations. We model queue cancellations according to a predefined input distribution (e.g., from the back of the queue, from the front of the queue, uniformly at random, etc.) Hence, passive orders are tracked according to historical LOB changes and  our cancellation assumptions and are executed whenever historical trades are. We assume that aggressive trades are always filled for the amount of liquidity available at the top level.

\begin{figure}[ht]
\vskip 0.2in
\begin{center}
\centerline{\includegraphics[width=\columnwidth]{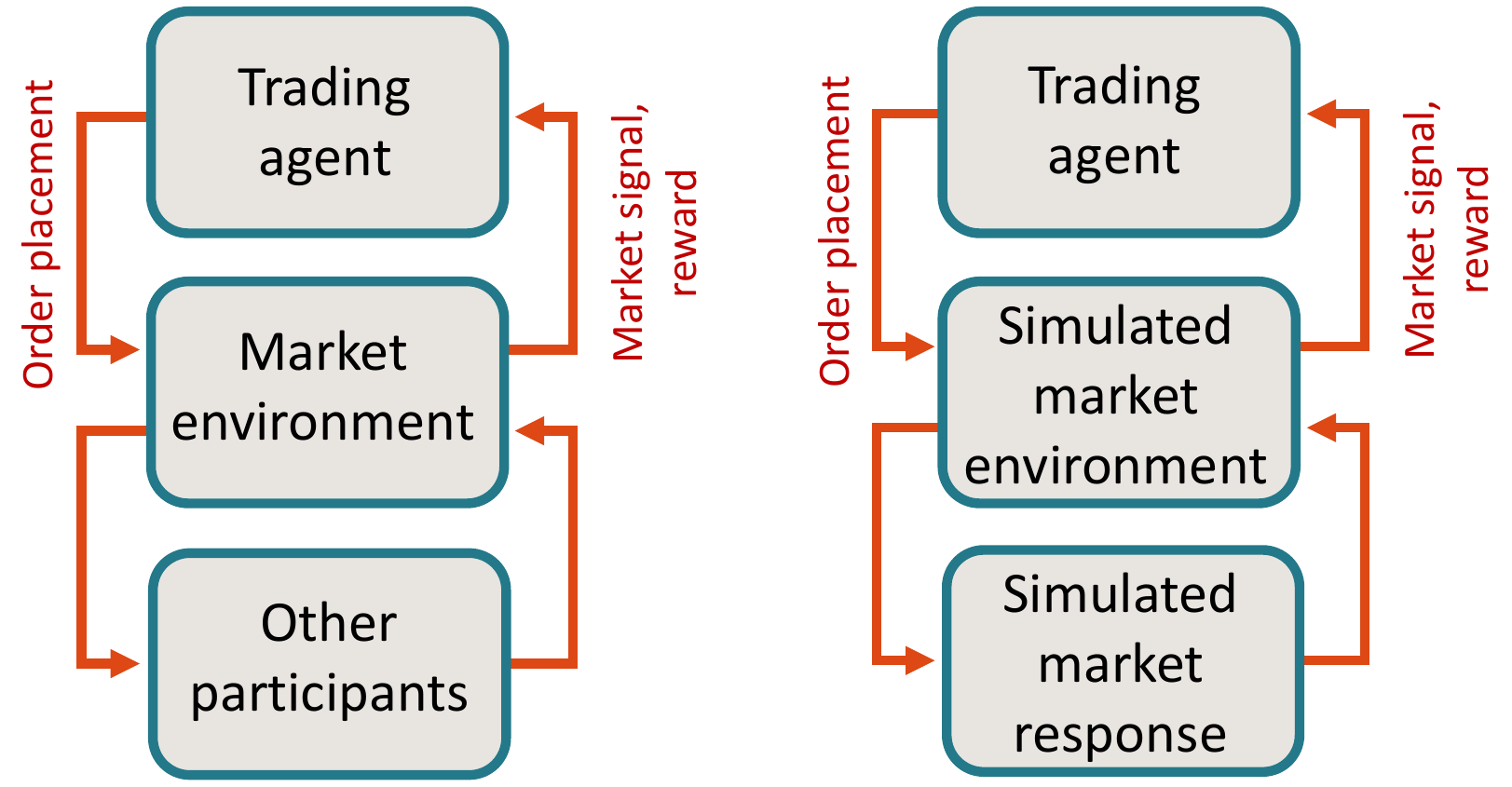}}
\caption{Schematic depiction of simulated vs real-time execution agents. }
\label{sim_vs_realtime}
\end{center}
\vskip -0.2in
\end{figure}

We emphasize that the autonomous agent can choose to place an aggressive order at any time. In real-time execution, aggressive orders will typically move the market.  However, when a discretionary aggressive action is executed in the simulated market environment, market response to it is not contained in the historical data. Therefore, in order for historical data replay in the simulated environment to be realistic, we need to synthesize the market response to our agent's actions (difference illustrated in Figure \ref{sim_vs_realtime}). Currently, we assume that there is no market impact to joining the queue passively. Simulating market response to our aggressive trading is described in detail in Section~\ref{market impact}.

Additionally, we model  latency between the agent's placement decisions and the time these decisions reach exchange. In practice, market latency depends on time of day (e.g. latency is typically higher at market open and close), and hence we use historical latency profiles to make relevant latency distribution assumptions.

To summarize, one can view LOB simulator as a multi-agent environment that models market response to autonomous agent's decisions given certain assumptions on the matching rules, queue cancellations and latency, which altogether define a set of simulation parameters $\psi$. We choose $\psi$ so that simulation output resembles real execution data as closely as possible. 

More formally, let $\setC$ be the sample space of execution costs of a given execution strategy over a selected historical time period. Define $\mathbf{P}(c),\ c \in \setC$, as the probability distribution determined by real-time trading; define $\mathbf{\tilde{P}}_\psi(c)$ as the probability distribution determined by trading simulated at random times. Then we can experimentally find a set of simulator parameters $\psi^*$ such that Kullback-Leibler divergence between $\mathbf{P}(c)$ and  $\mathbf{\tilde{P}}_\psi(c)$  is minimized:
\begin{equation}
\label{divergence}
\psi^* = \argmin{\psi} \sum_{c \in \setC} \mathbf{P}(c) \ln \frac{\mathbf{P}(c)}{\mathbf{\tilde{P}}_\psi(c)}.
\end{equation}

\subsection{Market impact}
\label{market impact}
Suppose at time $t$ our simulated autonomous trading agent decides to execute a child order of size $o_t$ aggressively. Let $v_t$ be the total volume available at the top LOB level opposite to our execution agent's side at time $t$.  We model market response to our aggressive action as a function of LOB liquidity and aggressive order size. In particular, we make the following observations about real-time trading price dynamics:
\begin{enumerate}
\item Adverse price move is a likely response to our agent's aggressive action.
\item Probability of adverse price move at tick $t'$ immediately following an aggressive action at tick $t$ increases as $o_t$ increases.
\item Probability of adverse price move at tick $t'$ immediately following an aggressive action at tick $t$ decreases as $v_t$ increases.
\end{enumerate}

Since aggressive action at time $t$ happened at our agent's discretion, market response to it was not included in historical LOB data. Therefore, simulated LOB time series needs to diverge from historical LOB time series in order to synthesize our agent's interaction with the market. Let $Y_t$ define the vector of historical LOB prices and volumes at multiple levels at time $t$. Similarly, define $\hat{Y}_t$  to be a vector of simulated  LOB prices and volumes at multiple levels at time $t$.  Also, define $t^*>t$ to be the time when next aggressive historical trade in agent's direction or next historical price move in the opposite direction occurs after our agent's aggressive action.

We model market response to our aggressive action at the next time step $t'$ post aggressive action at time $t$  as
\begin{align}
\label{no_impact}
\hat{Y}_{t'} =  Y_{t'} & \text{ , if   }  \frac{v_t}{o_t} > C_{MI} \\
\label{impact} 
\hat{Y}_{t'} = Y_{t^*} & \text{ , if   } \frac{v_t}{o_t}  \le C_{MI}, 
\end{align}
where $C_{MI}>0$ is a simulation parameter that defines the threshold below which order of size $o_t$ will move the market at time $t'$ \cite{Bouchaud17}. Effectively, (\ref{impact}) says that if $ \frac{v_t}{o_t} \le C_{MI}$, then the simulated time series $\hat{Y}_t$ will diverge from historical time series $Y_t$ until the next historical aggressive trade in agent's direction or historical market move in adverse direction happens. $C_{MI} \in \psi$ can be obtained using the simulator tuning procedure as described in (\ref{divergence}). The market will then return to its equilibrium and our trading impact will diminish in line with the historical data.

\section{Agent training}
\label{agent training}

In this section we describe how to build an autonomous trading agent given by a compact decision tree that will interact with the market to minimize cumulative execution cost subject to constraint on risk appetite.

\subsection{MDP formulation of order execution}
The order execution process can be formulated as an infinite-horizon MDP with a self-absorbing terminal state. At time $t_\sigma$, an autonomous agent receives a task to execute a large parent order of size $O_{t_\sigma}$ in the market. For time $t>t_\sigma$ the agent observes state variables that describe both its own status and the market status. Any change in state variables will trigger one step in the MDP, leading the agent to take an action of placing a child order of size $o <<  O_{t_\sigma}$ either passively or aggressively. The process continues until the order is liquidated. State evolution triggered by changes in observable variables follows our tick-based market simulator, and transition probabilities from one state to another under taking certain actions are implied by the historical LOB simulator with market impact synthesis as described in Section~\ref{simulator}. As in \cite{Nevmyvaka06}, the Markovian assumption in MDP formulation implies that the optimal action at any given state is independent of the agent's previous execution actions. 

We evaluate the agent performance by calculating its execution yield, which is defined as the relative performance of our sequential RL-based order execution compared to execution of the the entire parent order $O_{t_\sigma}$ at the mid price $p_\sigma$ at its arrival time. By setting an execution yield  of each individual fill as a reward, we train the agent to maximize its risk-adjusted cumulative rewards.

\textit{\textbf{States:}} We design the state space $\setS$ to include both a trading agent state variable and market environment variables. The agent state variable describes a position that is remaining for the agent to liquidate at time $t$. Note that the state with remaining position 0 is a self-absorbing terminal state in our infinite-horizon MDP. The market environment variables describe the LOB state and can be derived from various LOB information known to the agent at time $t$ such as bid-ask spread, volumes at multiple LOB levels, price offsets and volatility as well as available information about LOB states of correlated assets. Both the agent and the market environment variables are discretized into bins. If each state variable $X_1, X_2, \ldots, X_{|\setS|}$ is discretized to $n_1, n_2, \ldots, n_{|\setS|}$ bins, then the total number of states is $|\setS| = n_1\times n_2\times\cdots\times n_{|\setS|}$.

\textit{\textbf{Actions:}} At each MDP step, the agent makes a trading decision based on the observed current state. In particular, the agent chooses between two actions: to place a child order of size $o$ passively (via a limit order) or aggressively (via a market order). As described in Section~\ref{assumptions}, if an order is placed aggressively, it will be matched immediately for the amount of liquidity available at the top LOB level. If an order is placed passively and assumes a position in the LOB queue, it may be executed at a later time fully or partially or may not be executed at all if the market moves away. It is likely that market state variables change before a pending passive order is filled. In this case, if price did not move and passive placement is chosen, then no action is taken at all. Otherwise, if price moved, and new action puts passive placement at a new price level, or in case of aggressive placement regardless of the price move, an existing pending passive order is canceled.

\textit{\textbf{Rewards: }} The agent receives a reward as a reinforcement signal every time when a child order is executed (partially or fully) in the market. At time $t$, if our pending passive or aggressive order is matched with the opposite interest in the market, we say that a fill of size $f_t$ at price $p_t$ is produced.
We define the reward $R_t$ as our order execution yield:
\begin{equation}
R_t = 
\begin{cases}            
(p_t - p_\sigma) \times f_t & \text{ sell\ order} \\
(p_\sigma - p_t) \times f_t & \text{ buy\ order}
\end{cases}
\end{equation}
With the immediate reward defined as above, the undiscounted cumulative rewards equal to the total execution yield of parent-level order execution. Therefore, RL algorithms that seek to maximize expected total rewards will learn policies that maximize the execution yield.

When an order is placed aggressively, we assume that the fill happens immediately following the action. When an order is placed passively, the fill may happen many steps after, resulting in delayed rewards (or may not happen at the placed price level at all). Hence, aggressive actions have high execution cost and low execution cost variance, whereas passive actions have low execution cost and high execution variance. We seek to achieve the trade-off by maximizing the risk-adjusted cumulative rewards, and rather than applying a utility function to the reward \cite{Neuneier97, Ritter17}, we use the risk-sensitive Q-learning framework \cite{Mihatsch02} described in the following section.

\subsection{Learning algorithm: risk sensitive Q-learning}
\label{sec:RS Q-learning}
Consider an infinite-horizon MDP with finite state space $\setS$, finite action space $\setA$, and $\gamma$-discounted rewards with $\gamma\in (0, 1)$. At each step $i$, the agent observes state $S_i\in\setS$ and selects an action $A_i\in\setA$ based on such observation. An immediate reward $R_i$ is then received. In the risk-neutral setting, the Q-value function $Q^\pi(s, a)$ is defined as the expected total discounted rewards starting from initial state $s$, initial action $a$, and following a policy $\pi$:
\begin{equation}
\label{eq:Q_pi}
Q^\pi(s, a) = \E_{\pi}\left[\sum_{i=0}^\infty \gamma^t R_i | S_0=s, A_0=a\right].
\end{equation}
The optimal risk-neutral policy $\pi^*$ seeks to maximize the expected total discounted rewards: 
\begin{equation}
\pi^* = \argmax{\pi}Q^\pi(s, a).
\end{equation}
The optimal Q-value function $Q^*(s,a) = Q^{\pi^*}(s,a)$ then satisfies the Bellman optimality equation:
\begin{equation}
\begin{split}
&d^* = R_i + \gamma\ \max_{a\in\setA}Q^*(S_{i+1}, a) - Q^*(S_i, A_i) \\
&\E_{S_{i+1}, R_i}\left[ d^* \right] = 0.  \label{eq:bellman}
\end{split}
\end{equation}
The Q-learning \cite{watkins1989learning} algorithm essentially runs a stochastic approximation on the Bellman optimality equation and updates the estimation of the optimal Q-value function as:
\begin{align}
&d_i =  R_i + \gamma\ \max_{a\in \setA} \hat{Q}(S_{i+1}, a) - \hat{Q}(S_i, A_i) \\
&\hat{Q}(S_i, A_i) \leftarrow \hat{Q}(S_i, A_i) + \alpha_i d_i
\end{align}
where $\alpha_i \in [0, 1)$ is the learning step size. Q-learning converges to a globally optimal policy when the learning step sizes that correspond to updating each state-action pair satisfy:
\begin{equation}  \label{eq:Q-convergence}
\sum^\infty_{i=0} \alpha_{i(s, a)} = \infty, \quad \sum^\infty_{i=0} \alpha^2_{i(s, a)} < \infty, \quad \forall s\in\setS, a\in\setA.
\end{equation}

In the risk-sensitive setting proposed in \cite{Mihatsch02}, a parameterized transform function $U^\beta(x)$ is constructed as:
\begin{equation}
U^\beta(x) =
\begin{cases} 
	(1-\beta)x & x > 0 \\
	(1+\beta)x & x \leq 0
\end{cases},\quad \beta \in (-1,1).
\end{equation}
By imposing this transform function onto the Bellman optimality equation, the optimal risk-sensitive Q-values $\underline{Q}^*(S_i, A_i)$ can be implicitly defined as the solution of:
\begin{equation}
\begin{split}
&\underline{d}^* = R_i + \gamma\ \max_{a\in\setA}\underline{Q}^*(S_{i+1}, a)- \underline{Q}^*(S_i, A_i)\\
&\E_{S_{i+1}, R_i}\left[U^\beta\left(\underline{d}^*\right)\right] = 0.  \label{eq:rs-bellman}
\end{split}
\end{equation} 
When $\beta=0$, equation (\ref{eq:rs-bellman}) reduces to the risk-neutral case in equation (\ref{eq:bellman}). When $\beta\rightarrow 1$, any one-step forward estimation that is better than the current estimation will be suppressed, resulting in $\underline{Q^*}$ being optimized under the worst case criterion. The learned policy, therefore, demonstrates the risk-averse property. On the contrary, $\beta\rightarrow -1$ results in the risk-seeking behavior. Hence, changing $\beta$ allows us to search policy that matches our risk preference.

Similar to Q-learning in the risk-neutral setting, the update step of risk sensitive Q-learning performs stochastic approximation on the modified Bellman optimality equation (\ref{eq:rs-bellman}):
\begin{align}
&\underline{d_i} =  R_i + \gamma\ \max_{a\in \setA} \underline{\hat{Q}}(S_{i+1}, a) - \underline{\hat{Q}}(S_i, A_i) \label{eq:rs-td}\\
&\underline{\hat{Q}}(S_i, A_i) \leftarrow \underline{\hat{Q}}(S_i, A_i) + \alpha_i U^\beta(d_i) \label{eq:rs-qupdate}.
\end{align}
Risk-sensitive Q-learning converges under the same conditions (\ref{eq:Q-convergence}) as risk-neural Q-learning does. Furthermore, the above risk-sensitive transformations can be incorporated into a wide range of other RL algorithms that require function approximation.

We apply the above defined risk-sensitive Q learning with tabular state-action presentation to to train our trading agent. The risk-sensitive Q-values are iteratively updated as suggested by equations (\ref{eq:rs-td}, \ref{eq:rs-qupdate}). As the rewards of passive orders are often delayed, we incorporate eligibility trace \cite{sutton1985temporal, precup2000eligibility} into the risk-sensitive RL framework to accelerate learning. Equation (\ref{eq:rs-qupdate}) updates the Q-values using one-step look forward, while the eligibility trace propagates back the received rewards and updates the visited states along the history. We also emphasize that once the trading agent is trained in the proposed simulator, the Q-policy can be continuously updated online as a result of real-time trading activities.

Generation of decision trees from a tabular policy is discussed in Section~\ref{decision_tree_example}.

\subsection{Feature selection via least-squares policy iteration}
\label{feature selection}

Function approximation for state-action values allows us to consider RL execution agent design over large state spaces. As a first step, we intend to select a small subset of statistically significant state variables from a large set of input state variables. The state variables are evaluated over a modified LOB dataset that includes market response to our aggressive actions in risk-neutral case. Then, as a second step, we discretize the selected state variables and use the risk-sensitive Q-learning framework defined in Section~\ref{sec:RS Q-learning} to train a risk-sensitive execution agent. In the final step, the learned policy is represented as a decision tree.

We approximate the $Q$-value function $Q^\pi(s, a)$ in (\ref{eq:Q_pi}) with a linear parametric function approximator, i.e. a linear weighted combination of $k$ state-action basis functions $\phi_j(s,a)$, $j=1,2,\ldots,k$:
\begin{equation}
Q^\pi(s, a) \approx  \tilde{Q}^\pi(s, a, w) = \vphi(s,a)^\top \vw = \sum_{j=1}^k \phi_j(s,a) w_j,
\end{equation}
where $w \in \mathbb{R}^k$ is a  parameter that we have to learn.
We then solve for a regularized solution to the sample-based linear system. We simulate a large number of order executions in an environment that synthesizes market response to our trading, and record total $N$ transition steps in the MDP. Define $\mPhi$ (respectively $\mPhi'$ ) $\in \mathbb{R}^{N \times k}$ to be the empirical state-action feature matrices whose rows contain state-action feature pairs $\vphi(S_i, A_i)^\top$ (respectively $\vphi(S_{i+1}, \pi(S_{i+1}))^\top$), and $\tilde{\vr} \in \mathbb{R}^N$ to be the reward vector consists of $R_i$ \cite{Lagoudakis03}. As in standard least squares temporal difference algorithm (LSTD) \cite{bradtke97}, we need to solve linear system

\begin{equation}
\tilde{\mA} \vw = \tilde {\vb},
\end{equation}
where
\begin{align}
\tilde{\mA} & =\tilde{ \mPhi }^\top \left( \tilde{\mPhi} - \gamma \tilde{\mPhi}' \right)  \\
\tilde{\vb} & = \tilde{\mPhi}^\top \tilde{\vr}.
\end{align}

As a regularization procedure, we use Dantzig-LSTD approach \cite{Geist12}:
\begin{equation}
\vw_{\lambda} = \argmin{\vw \in \mathbb{R}^k} \lVert{\vw\rVert}_1 \quad
\text{subject to } \lVert{\tilde{\mA}\vw - \tilde{\vb}\rVert} \le \lambda  
\end{equation}
for $\lambda > 0$, which can be solved efficiently using linear programming.

We collect a large number of simulated execution samples, solve for regularized solution of resulting linear system and then use model-free least squares policy iteration (LSPI) \cite{Lagoudakis03} to iteratively improve over the regularized policy. We note that, as with any approximate policy iteration algorithm, convergence of LSPI is not guaranteed.  Furthermore, it is known that LSTD computes an approximation that is weighed by the state visitation frequencies and can oscillate between bad policies \cite{Koller2000}. In practice, we are however able to use the above framework to select a small number of state-action pairs that we further use for decision tree generation using risk-sensitive Q-learning. 

\section{Results}
We train a risk-sensitive decision tree execution agent as described in Section ~\ref{sec:RS Q-learning} using high frequency futures data. The future that we train our strategy for is considered a liquid asset. Both historical tick LOB data and completed trade data are public and provided by the exchange.

\subsection{Risk-sensitive strategy selection}

\begin{figure}[ht]
\begin{center}
\centerline{\includegraphics[width=\columnwidth]{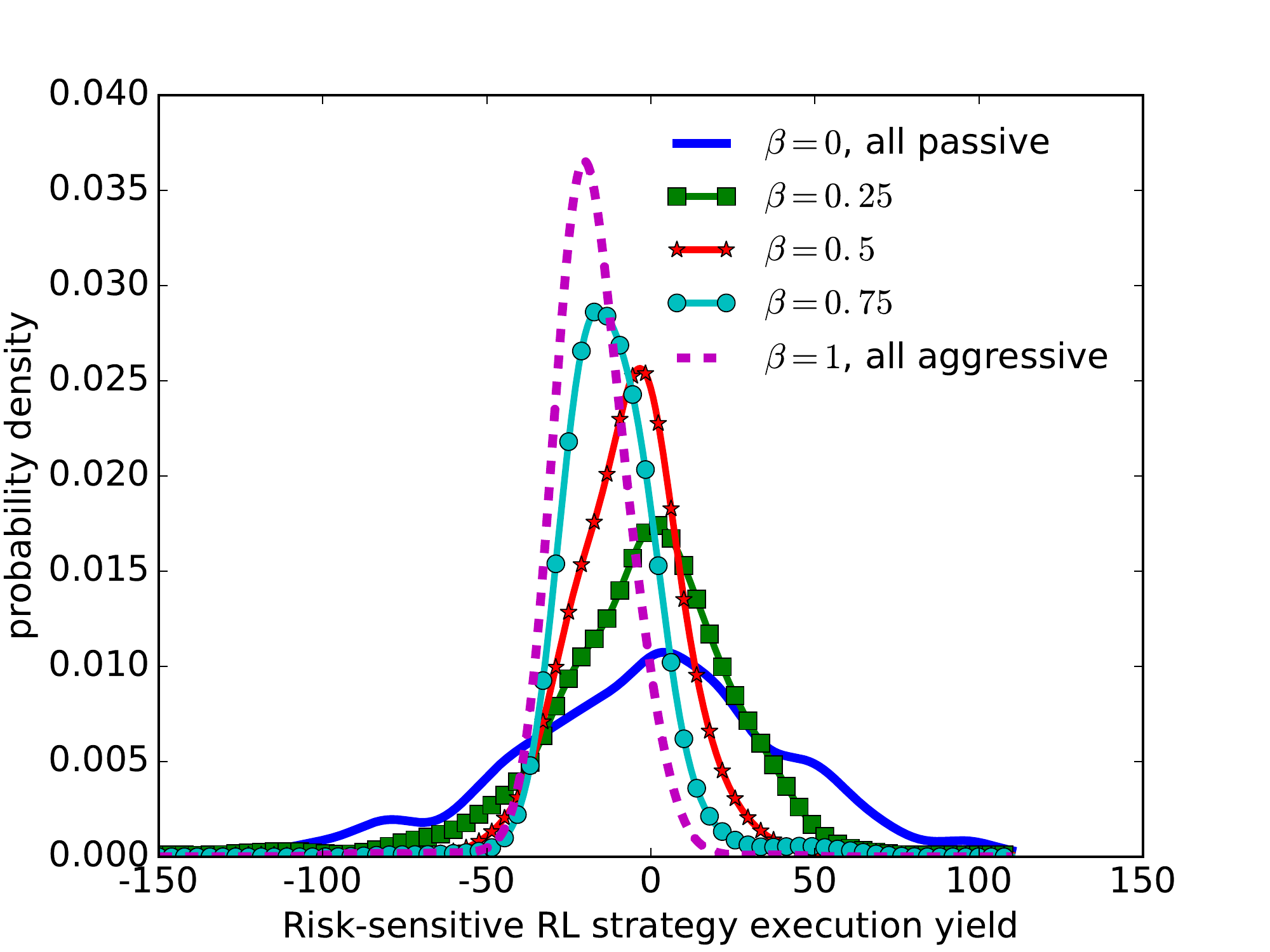}}
\caption{Distribution of execution yield under policies learned by risk-sensitive Q-learning with different risk-sensitivity parameters $\beta$.}
\label{fig:choose_beta}
\end{center}
\vskip -0.2in
\end{figure}

For a chosen set of state variables, we ran the risk-sensitive Q-learning algorithm described in Section~\ref{sec:RS Q-learning} with different risk-sensitivity parameter $\beta$ ranging from $0$ to $1$. The distribution of  resulting execution yield under corresponding learned policies is shown in Figure (\ref{fig:choose_beta}). The case of $\beta=0$ resulted in all passive order executions. It can be explained by the fact that the expected cumulative reward of passive executions was the highest, since bid-ask spread was never crossed. However, the variance was also the highest due to LOB price moves away from the agent while passive orders were pending in the queue. On the contrary, $\beta=1$ resulted in all aggressive order placements, which had the lowest expected cumulative reward as the bid-ask spread was always paid, and the lowest variance as all child orders were executed momentarily by our simulator assumptions. While changing $\beta$ from $0$ to $1$, we observed the decrease in both expectation and variance, indicating a move from all-passive to all-aggressive policy. Therefore, we observed that $\beta$ selection procedure is well aligned with the theoretical behavior of risk-sensitive RL described in Section \ref{sec:RS Q-learning}. An acceptable range of $\beta$ is then chosen based on trader's risk appetite.

\subsection{RL execution agent presentation by decision trees}
\label{decision_tree_example}

For a given parameter $\beta$ and a given set of state variables, we train a tabular Q-learning policy using the procedure described in Section~\ref{sec:RS Q-learning} with discretized state space. A tabular policy maps state variables to actions, which allows us to represent a learned tabular execution agent by a decision tree. Decision boundaries are implied from the state discretization. 

\begin{figure}[ht]
\vskip 0.2in
\begin{center}
\centerline{\includegraphics[width=\columnwidth]{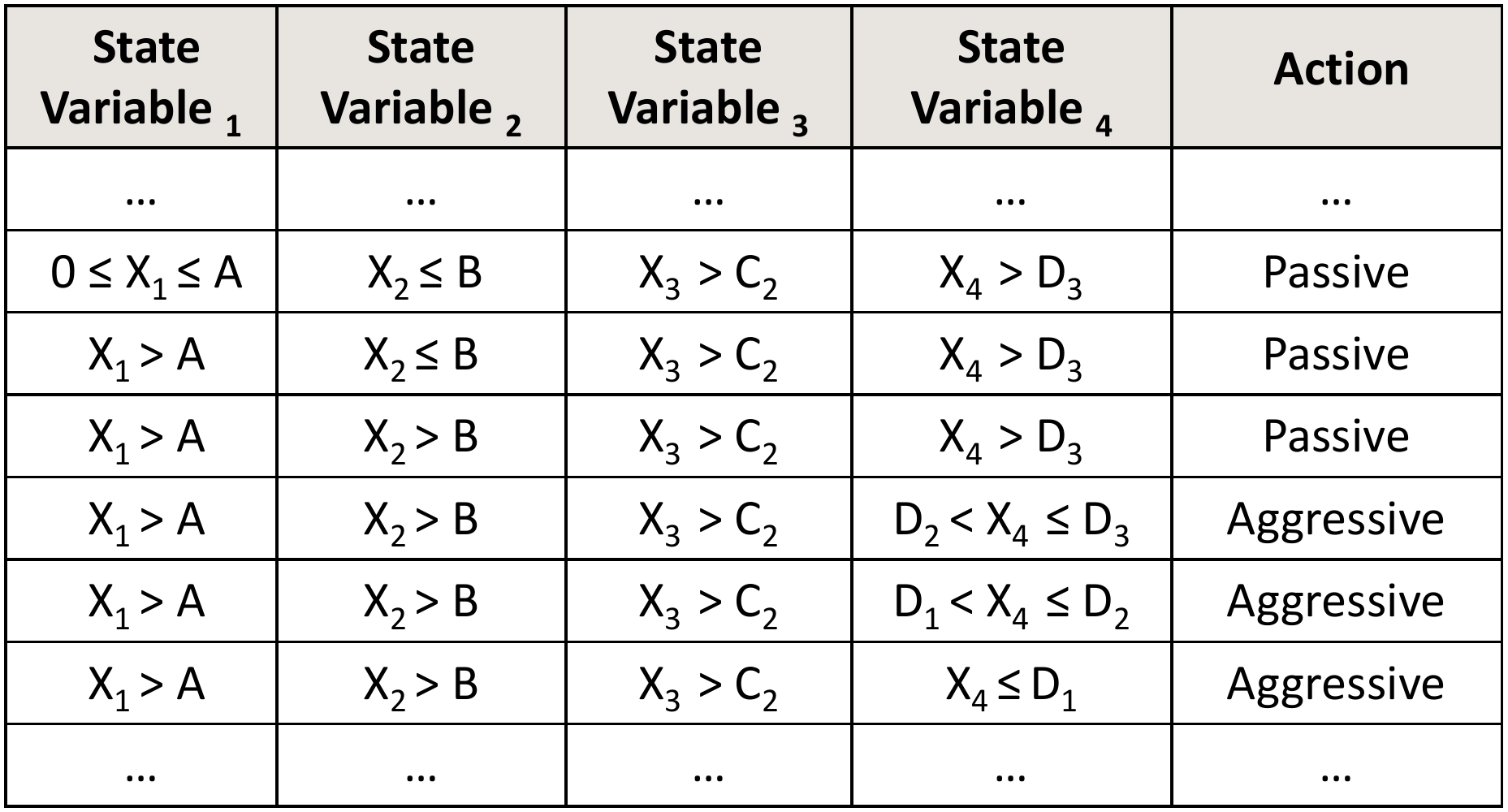}}
\caption{ Fragment of a tabular optimized execution policy.  }
\label{dec_tree_table}
\end{center}
\vskip -0.2in
\end{figure}

\begin{figure}[ht]
\vskip 0.2in
\begin{center}
\centerline{\includegraphics[width=\columnwidth]{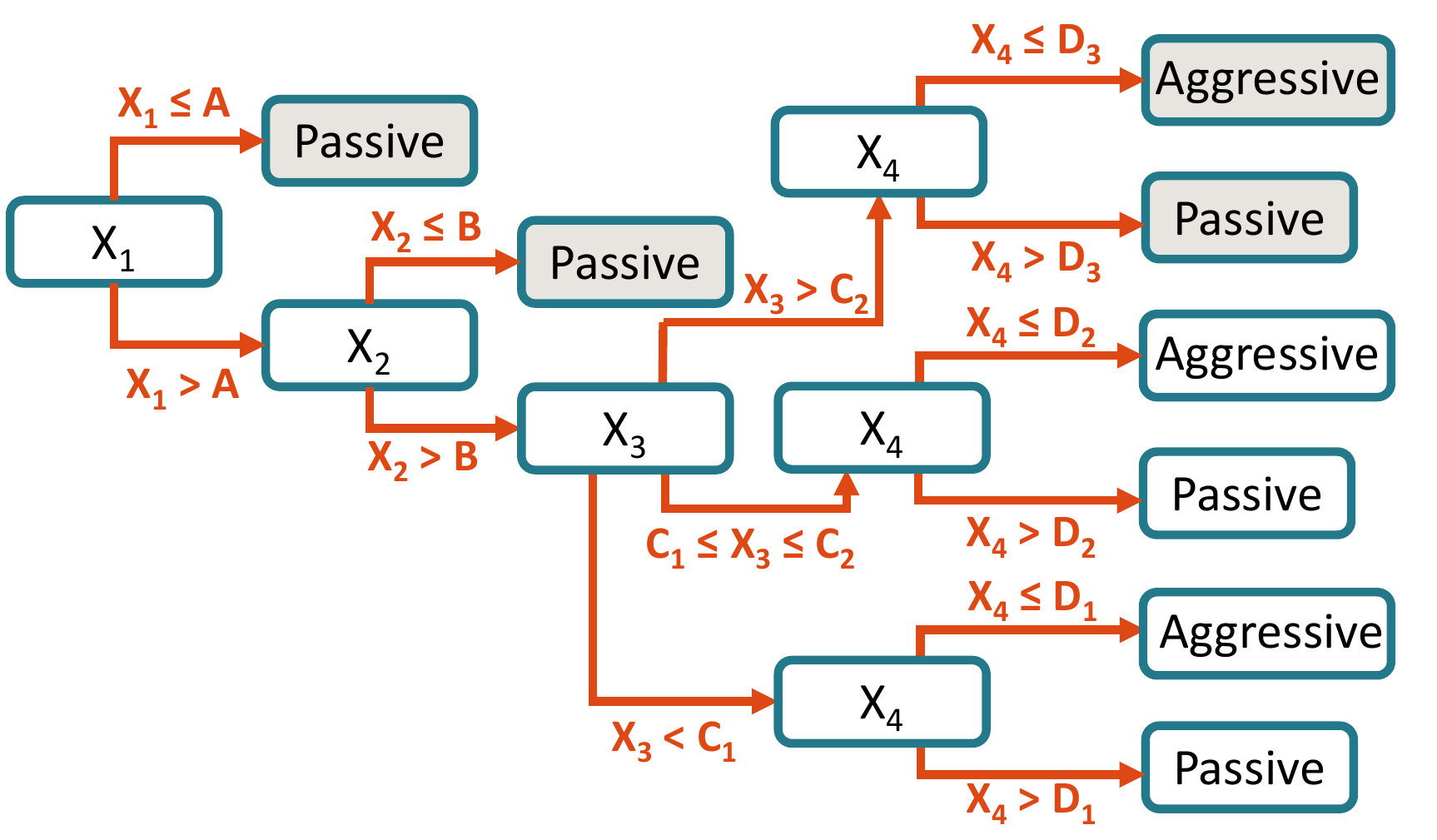}}
\caption{Structure of an execution agent given by a decision tree. Rows of a tabular policy that correspond to the highlighted action nodes are given in  Figure~\ref{dec_tree_table}.}
\label{decision_tree}
\end{center}
\vskip -0.2in
\end{figure}

For example, consider an MDP with one agent and three environment state variables, where $X_1$ is the agent state variable and $X_2, X_3, X_4$ are the environment state variables. Empirically, we discretize the state variables into the following bins:
\begin{align*}
X_1 \in & [ 0; A ] \cup (A; +\infty) \\
X_2 \in & ( -\infty; B] \cup ( B; +\infty) \\
X_3 \in & ( -\infty; C_1] \cup (C_1; C_2] \cup ( C_2; +\infty)\\
X_4 \in & ( -\infty; D_1] \cup (D_1; D_2] \cup (D_2; D_3] \cup ( D_3; +\infty),
\end{align*}
where  $A$, $B$, $C_1$, $C_2$, $D_1$, $D_2$, $D_3 \in \mathbb{R}$ are chosen so that the number of RL training samples in each bin is of the same order. Figure~\ref{dec_tree_table} shows the fragment of a tabular policy that is subsequently processed into the decision tree given in Figure~\ref{decision_tree} by feeding its rows to the classic Hunt's algorithm \cite{hunt66}, which determines the exact topology of a decision tree.

\subsection{Comparison to benchmark}

\begin{figure}[ht]
\vskip 0.2in
\begin{center}
\centerline{\includegraphics[width=\columnwidth]{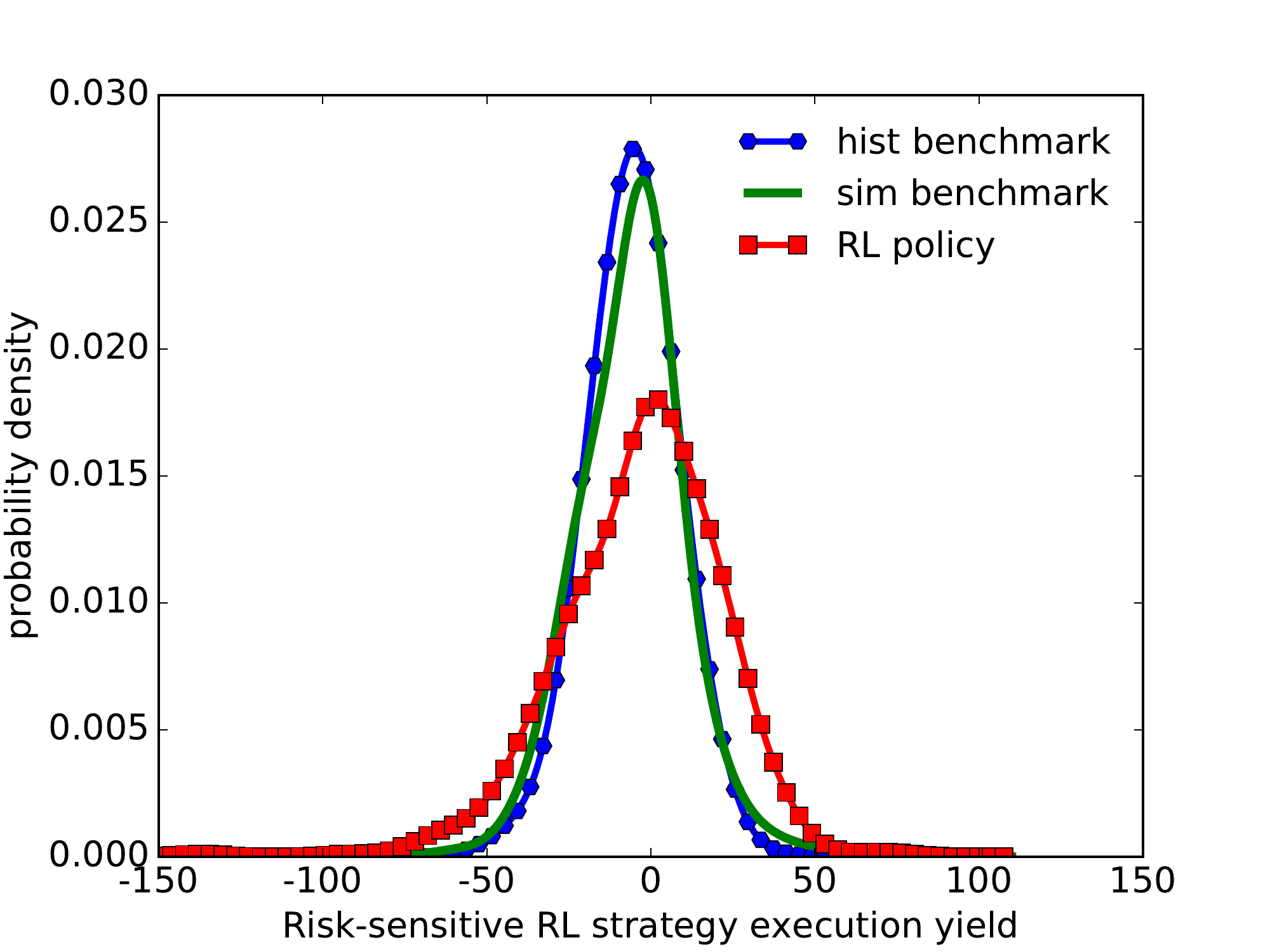}}
\caption{Performance of the RL decision tree execution agent compared to historical non-RL benchmark strategy.}
\label{sim_agent}
\end{center}
\vskip -0.2in
\end{figure}

As described in Section~\ref{assumptions}, we first select a set of parameters $\psi$  so that the divergence between the historical benchmark strategy (non-RL based) and the same simulated benchmark strategy yields (\ref{divergence}) is minimized. This ensures that the simulator assumptions are aligned with historical benchmark executions. Then we train a decision tree reinforcement learning strategy inside the simulated environment for a given parameter $\beta$ and a given set of state variables. Figure~\ref{sim_agent} illustrates the above process: the RL decision tree execution agent achieves $32\%$ in execution cost savings at the expense of $27\%$ increase in standard deviation of the cost.

\section{Conclusions and future work}
We showed a framework for risk-sensitive RL-based decision tree autonomous agent training. In particular, the framework that we presented consists of a LOB simulator that is capable of synthesizing market response to aggressive trading from historical data, and a risk-sensitive Q-learning procedure that can produce compact execution agents given by decision trees.

In algorithmic trading, actions of any given agent incur response from other market participants. Agent-based modeling and simulation is a relatively new approach that can potentially allow to simulate interaction between individual market participants \cite{macal10}. We also would like to underline the importance of training reinforcement learning models on synthetic data for trading execution applications.  In particular, viewing market response to our aggressive trading given by (\ref{no_impact}) and (\ref{impact}) as a reinforcement learning policy and trying to learn its parameters can be instrumental to achieving the goal of synthesizing market impact by the online RL training \cite{Ruiz19}.

Decision trees are a practical representation of trading agents. In the future, we would  like to explore the use of deep learning as a method to capture complex dependencies from rich feature set (time dependencies in particular), and to explore reduction of neural network architectures to compact decision trees with a small number of most significant states (example of similar recent work on neural network reduction to decision trees is given in \cite{Frosst17}). 

Use of deep neural networks for value function approximation is the key idea in the recent development of designing agents for playing the games of Atari and Go \cite{mnih2013playing, Silver_2016}. It has achieved impressive results and inspired other uses of deep neural networks for agent representation, including deep neural networks agents for trading applications as they can be well suited to capture the non-linear underlying price dynamics. Among others, a deep Q-trading system was presented in \cite{wang2016deep}. LSTM presentation of RL trading agent was given in \cite{Lu17}. Hedging a portfolio of derivatives (including over-the-counter derivatives) in the presence of market frictions was considered in \cite{Buehler18}, where deep RL was applied to non-linear reward structures. In the future, we would like to compare  performance of decision tree execution agents to agents given by neural networks and recurrent neural networks (LSTM in particular) in order to better understand the benefits of each.

\section{Acknowledgments}
Authors would like to thank Mark Rubery for helpful discussions, as well as Tucker Balch and Shilong Yang for useful comments on the manuscript.

\bibliography{MAL_ICML_decision_tree}
\bibliographystyle{icml2019.bst}

\end{document}